\documentclass{PoS}
\pdfoutput=1

\title{Development of Slow Control Boards for the Large Size Telescopes of the Cherenkov Telescope Array}

\ShortTitle{Development of Slow Control Boards for LSTs of the CTA}

\author{\speaker{Daniela Hadasch} \\
        Institute for Cosmic Ray Research, The University of Tokyo\\
        E-mail: \email{hadasch@icrr.u-tokyo.ac.jp}}

\author{Yusuke Konno$^{a}$, Hidetoshi Kubo$^{a}$, Daisuke Nakajima$^{b}$, Hideyuki Ohoka$^{b}$, Takayuki Saito$^{a}$, Masahiro Teshima$^{b}$
     {and the LST team for the CTA consortium}\\
     \llap{$^a$} Kyoto University, Japan\\
     Kitashirakawa, Sakyo, Kyoto, 606-8502, Japan\\
     \llap{$^b$} Institute for Cosmic Ray Research, The University of Tokyo, Japan\\
     5-1-5, Kashiwanoha, Kashiwa-city, Chiba, 277-8582, Japan\\
     E-mail:  \email{konno@cr.scphys.kyoto-u.ac.jp}, \email{kubo@cr.scphys.kyoto-u.ac.jp}, \email{dnakajim@icrr.u-tokyo.ac.jp}, \email{hohoka@icrr.u-tokyo.ac.jp}, \email{tysaito@cr.scphys.kyoto-u.ac.jp}, \email{mteshima@icrr.u-tokyo.ac.jp}}

\abstract{The camera of the Large Size Telescopes (LSTs) of the Cherenkov
Telescope Array (CTA) consists of 265 photosensor modules, each of them
containing 7 photomultiplier tubes (PMTs), a slow control board (SCB), a
readout board, and a trigger logic. We have developed the SCB, which is
installed between the 7 PMTs and the readout board. The main task for SCBs
is the controlling of the high voltages for the PMTs and the monitoring of their
anode currents. In addition, the SCB provides the functionality to create test pulses
that can be injected at the input of the PMT preamplifier in order to emulate a
PMT signal without the need of setting a high voltage, or even without the PMT
itself. The test pulses have a very similar width as the PMT pulses (less than
3\,ns FWHM) and their amplitude can be adjusted in a wide dynamic range. These
features allow us not only to test the functionality of the camera modules but
also to fully characterize these. We report on the design and the functions of
the SCB together with the results of test measurements.}

\FullConference{The 34th International Cosmic Ray Conference,\\
		30 July- 6 August, 2015\\
		The Hague, The Netherlands}

\begin{document}

\section{Introduction}

The next generation observatory for very high energy $\gamma$-rays will be the Cherenkov Telescope Array (CTA).
This new observatory is now in the planning and prototyping phase.
It will be built on two sites: one array will be constructed in the Northern, the other one in the Southern hemisphere.
Four LSTs (Large Size Telescope) of 23\,m diameter and 28\,m focal length will be arranged at the centre of both arrays to lower the energy threshold and to improve the sensitivity of CTA below 200\,GeV.
The first LST is being manufactured as a fully functional prototype that is installed directly on the site and becomes the first LST of CTA (Pre-Construction), once commission finishes and it has been verified that it fulfills all CTA requirements.
This prototype will be installed on the Canarian Island La Palma at the Observatory of El Roque de los Muchachos.

The camera of the LST is based on a modular design with all the electronics on-board, i.e. contained in the camera body. It is conceptually divided in three main parts:
\begin{itemize}
\item The Focal Plane Instrumentation (FPI).
\item Seven-pixel modules that contain the front-end electronics.
\item The global and auxiliary camera elements (for example power supplies, environmental control, reference LEDs, trigger interface board etc. for details see \cite{abc}).
\end{itemize}

\begin{figure}
\begin{center}
\includegraphics[width=.9\textwidth]{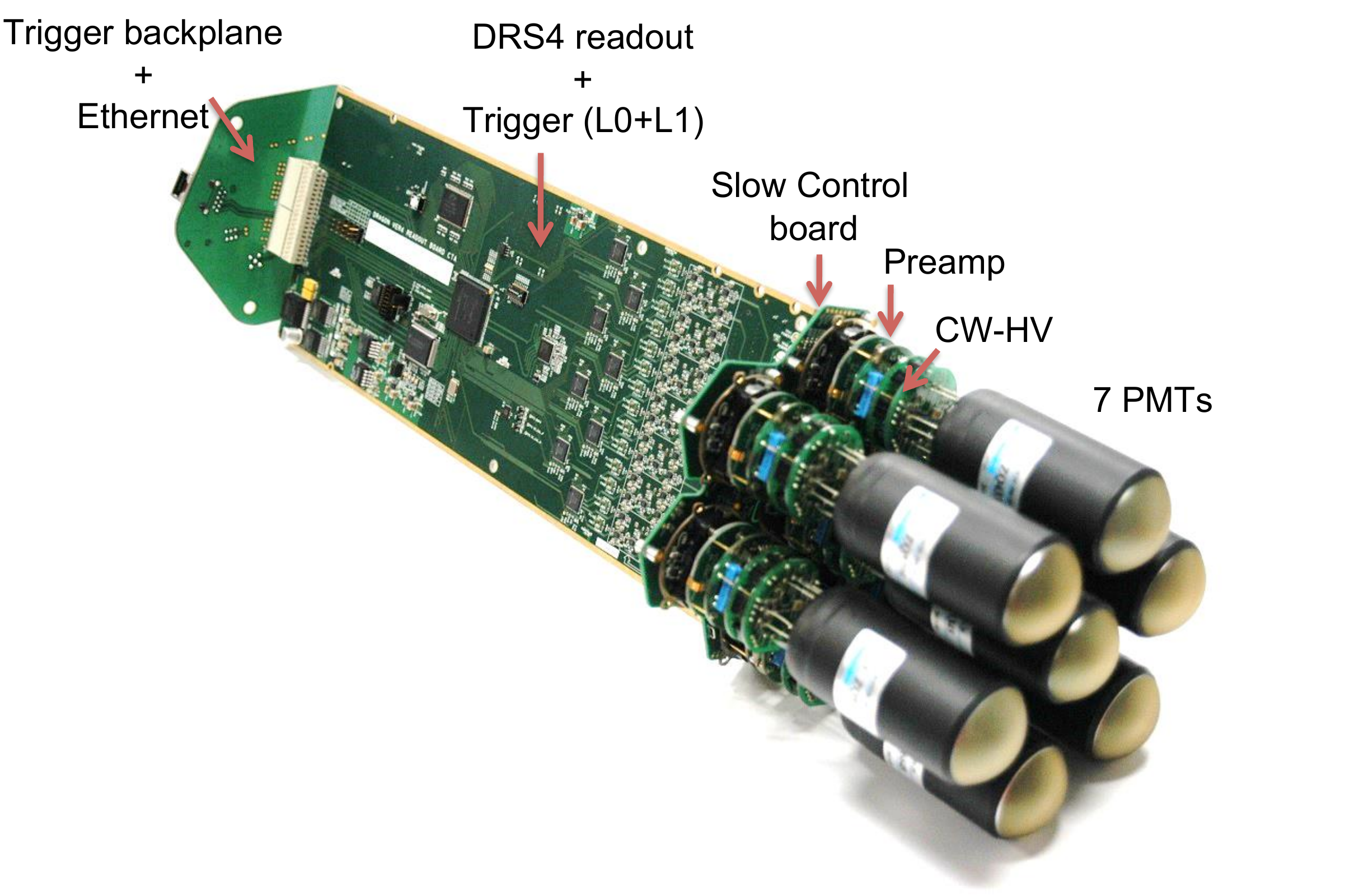}
\caption{Photo of a \textit{PMT} module. The core of the readout system is based on the Domino Ring Sampler chip (DRS4), which is an application specific integrated circuit (ASIC) of switched capacitor arrays (SCAs). For details see \cite{shu}.}
\label{module}
\end{center}
\end{figure}

The camera will consist of 265 photosensor modules, each of them containing 7 photomultiplier tubes (PMTs), a slow control board (SCB), a readout board (Dragon board), and a trigger logic.
Details on the Dragon board can be found in this contribution: \pos{862} \cite{shu}.
A photo of such a photosensor module with seven pixels is shown in Figure~\ref{module}.
Between the readout board and the PMTs the SCB is installed, which is developed by us and which will be described in the following sections.
Further technical details on the LST can be found in the Technical Design Report \cite{abc}.

\section{Slow Control Board}
\begin{figure}
\begin{center}
\includegraphics[width=.49\textwidth]{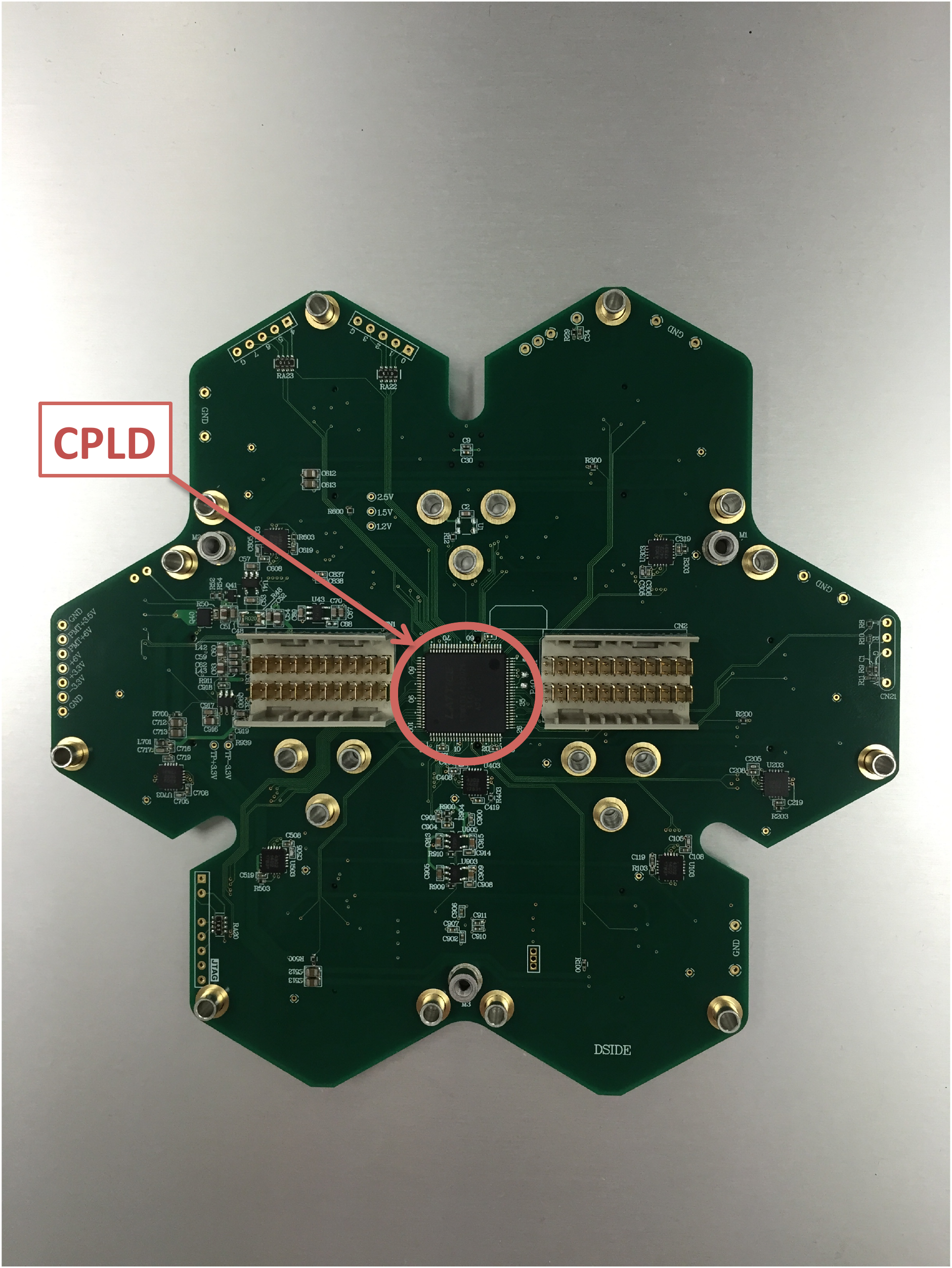}
\includegraphics[width=.49\textwidth]{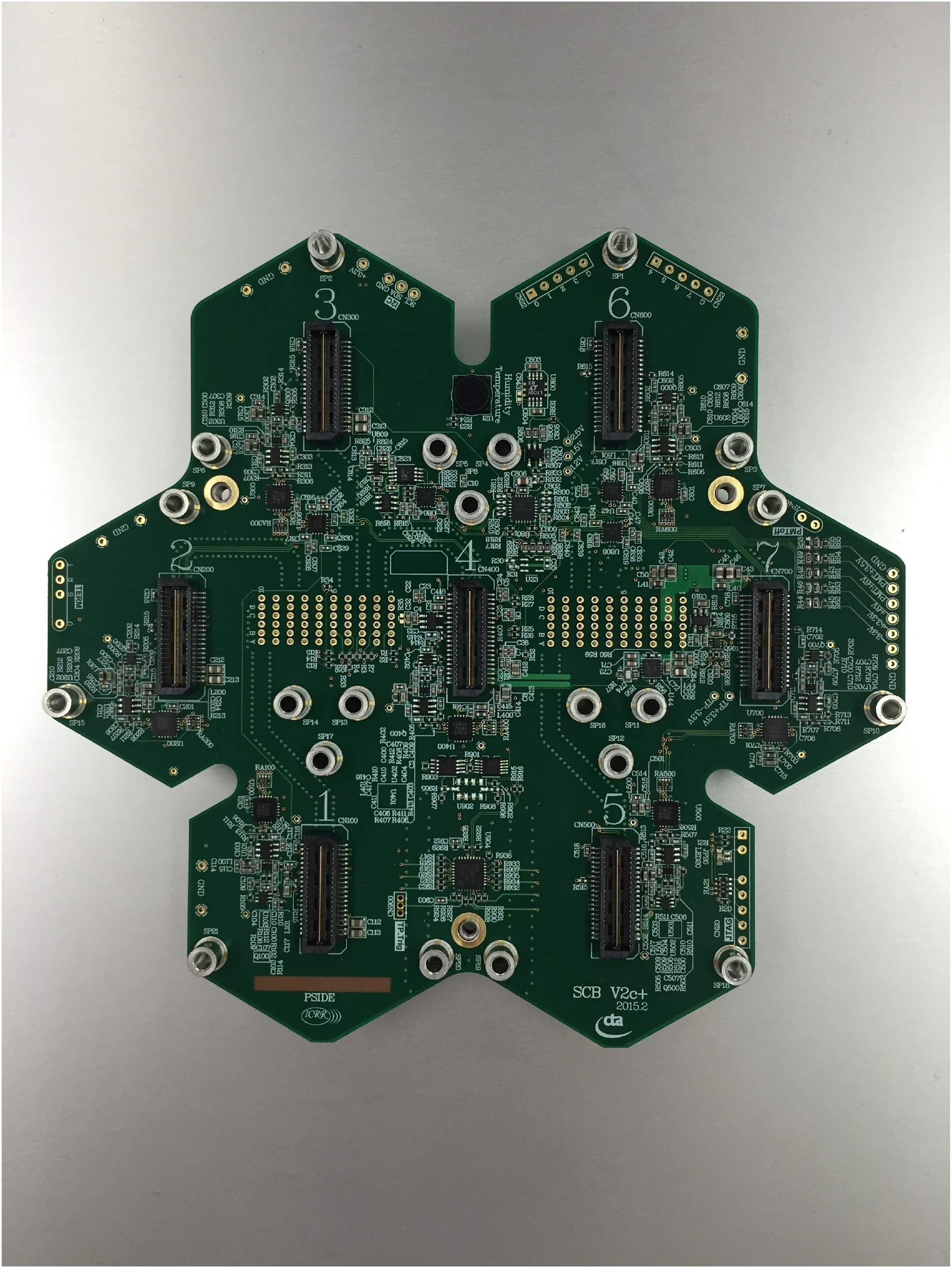}
\caption{\textit{Left:} Photo of the side of the SCB connected to the Dragon board. \textit{Right:} Photo of the side of the SCB connected to the PMTs.}
\label{scb}
\end{center}
\end{figure}

We developed the SCB control such that each PMT can be monitored
independently. The SCB controls the Cockcroft-Walton High Voltage (CW-HV) supply with
a Digital to Analogue Converter (DAC) and it monitors both CW-HV and PMT anode DC currents with an Analogue to Digital Converter (ADC). A temperature sensor, humidity sensor and a circuit for monitoring the power
voltage are also implemented. 
Figure~\ref{scb} shows a photograph of such a SCB.

The SCB has different functions: controlling the high voltages for the PMTs, monitoring their anode currents and temperatures and creation of test pulses.
To control these different functions on the SCB, a CPLD (Complex Programmable Logic Device)\footnote{The CPLD is indicated in Figure~\ref{scb} with a red circle.} is implemented in the SCB.
This CPLD communicates with the Field Programmable Gate Array (FPGA) of the readout electronics board via Serial Peripheral Interface (SPI) communication. 
Also the CPLD firmware is configurable via SPI communication. Seven PMT modules are attached to one side of
the SCB and the readout electronics board is connected at the other side. Since
the PMT signal is fed to the readout board through the SCB, the routing of PMT
signal has been carefully optimized to the same layout length. The voltages of a SCB are
$\pm$ 3.3\,V and 6\,V which are the same levels used for the analogue memory readout board. The
power consumption of one SCB is $\sim$22\,mW in operational mode and at most $\sim$1221\,mW during pulse injection, which will be described in the next sections.

\subsection{Test Pulse Operation}
The SCB can generate a narrow test pulse
that can be used as a test input even during the day. The pulse amplitude can
be adjusted with different attenuations and it simulates properly the typical
signal from a PMT. The pulse is injected into the pre-amplifier board and it
propagates along the whole electronic chain to all channels. 
Using this functionality, the performance of the Dragon boards can be carefully checked even after the installation in the camera.

\subsection{Pulse Shape}
\begin{figure}
\begin{center}
\includegraphics[width=.6\textwidth]{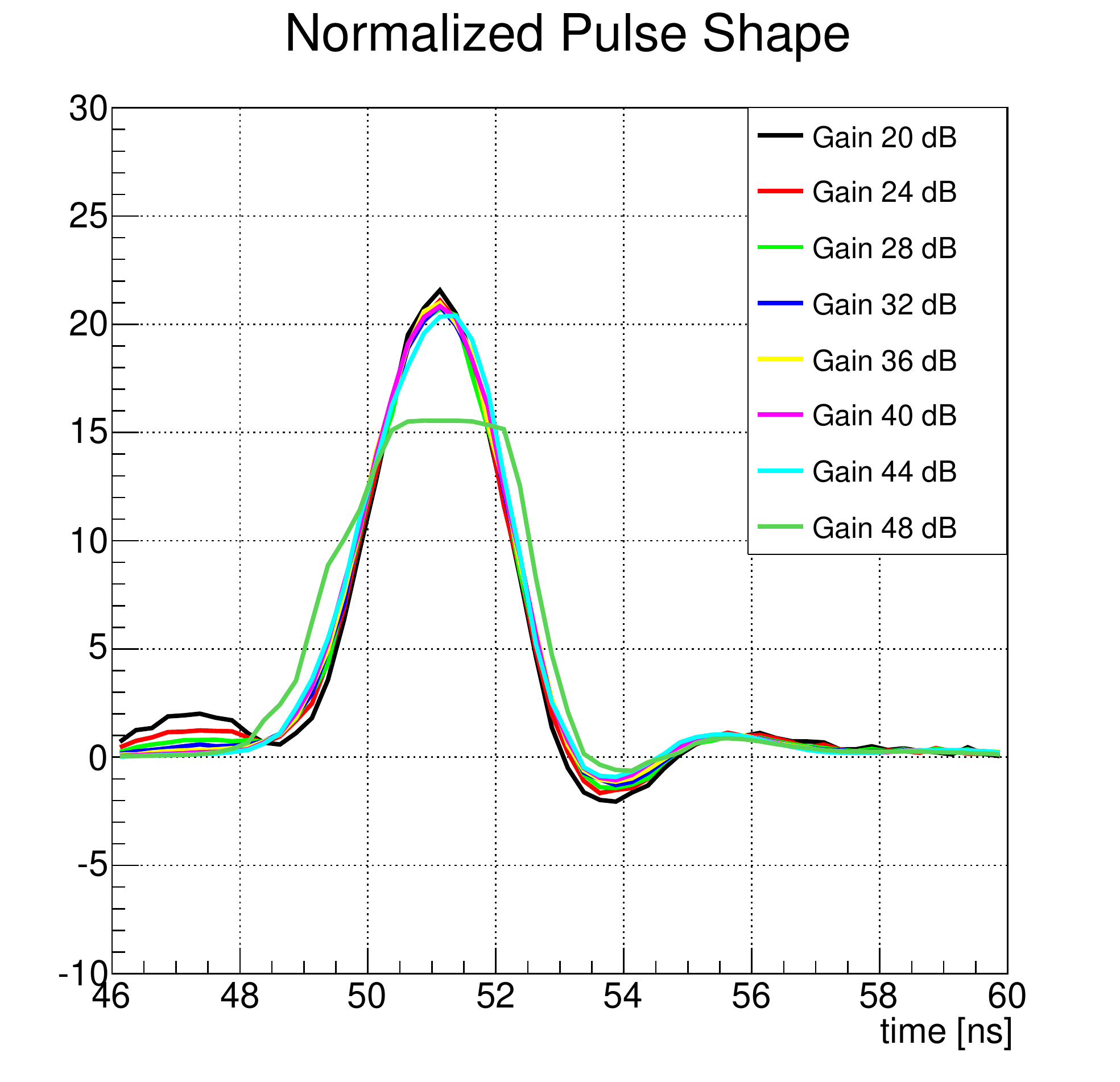}
\caption{Normalized shape of the narrow test pulses recorded by a Dragon board. For each gain, time corrected average of 300 pulses is shown.}
\label{shape}
\end{center}
\end{figure}

The FPGA on the Dragon board sends a LVDS pulse to the SCB. The width of
the pulse can vary between 7.5\,ns and a few microseconds. Receiving the signal,
the SCB  shapes it into a narrow pulse with a FWHM of 2.5\,ns, which is
similar to the width of the nominal pulses of the used PMTs. It is also possible to skip
this shaping. Afterwards, the pulse is distributed to 7 PMT channels. Each channel
has a variable attenuator and a "boosting switch". The attenuation factor of
the attenuator can vary by 64\,dB with a step size of 1\,dB. By activating the
boosting switch, the pulse amplitude increases by a factor of 10, providing
additional 20\,dB. Therefore, the dynamic range of the test pulse amplitude is more than 4
orders of magnitudes (84\,dB).
Figure~\ref{shape} shows the normalized shape of the short test pulses recorded
by a Dragon board. FWHM is about 2.5\,ns and no deformation of the shape is
seen for different amplitudes\footnote{The pulse for a gain of 48 shows a flat top due to saturation.}. This fast pulse is useful to test the linearity,
band width and cross-talk level of Dragon boards.

\begin{figure}
\begin{center}
\includegraphics[width=.6\textwidth]{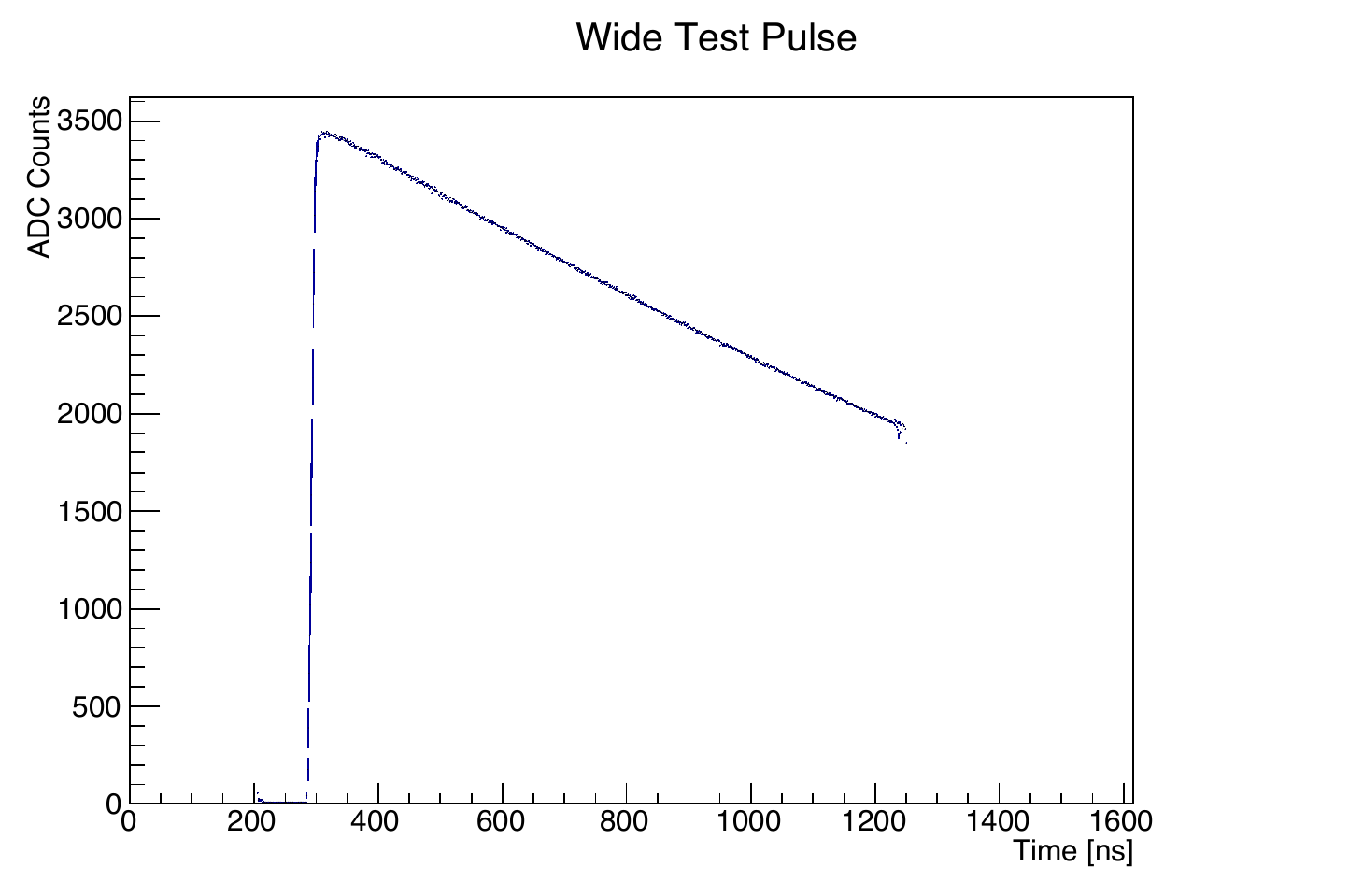}
\caption{Example of a wide test pulse.}
\label{wide}
\end{center}
\end{figure}

Figure~\ref{wide} shows an example of wide test pulse. It is produced by bypassing the shaping circuit of the SCB. This is also useful to check the performance of all capacitors in the DRS4 chip on the Dragon boards. The DRS4 chip has 8192 capacitors and from time to time some of them do not work. With this microsecond-wide pulse, one can easily find such bad capacitors.

\subsection{Dynamic Range}
\begin{figure}
\begin{center}
\includegraphics[width=.9\textwidth]{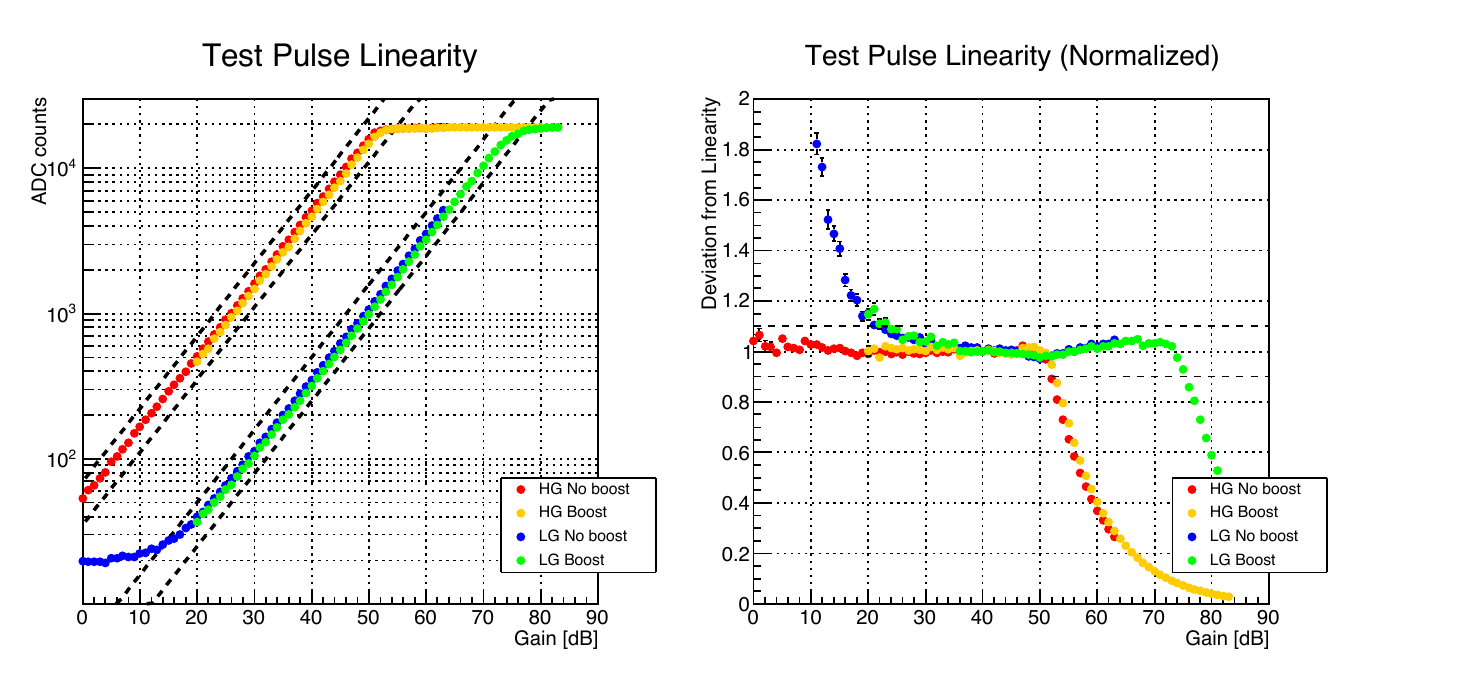}
\caption{Linearity plots. \textit{Left:} Extracted charge against the gain (inverse of the attenuation factor) of
the test pulse. \textit{Right:} Deviation from linearity. See text for details.}
\label{linearity}
\end{center}
\end{figure}

Using this SCB, the linearity of the Dragon board was measured. Short pulses
(2.5\,ns FWHM) were injected and the charge was extracted with a 3\,ns integration
window. The extracted charge against the gain (inverse of the attenuation factor) of
the test pulse is plotted in the left panel of Figure~\ref{linearity}.
Red and orange points indicate non-boosted and boosted test pulses injected in a high gain channel.  Blue and green denote signals injected in the low gain channel.

The right panel of Figure~\ref{linearity} shows the deviation from linearity. The deviation was computed in the following way

\begin{equation}
\Delta Q_{i} = \frac{Q_{i}}{Q_{40}} \frac{1}{10^{(i - 40)/20}}
\end{equation}
where $Q_i$ denotes the charge at a test pulse gain of "i".

The high gain channel shows linearity from 0\,dB to 50\,dB, which corresponds to 0.3 to 100 photoelectrons. After 50\,dB, the readout ADC saturates.
Below 20\,dB, the low gain channel is not linear. This is due to readout noise. If
the amplitude of the test pulse is as low as or lower than the noise level,
charge extraction fails, resulting in an overestimate of the charge. The low Gain
channel saturates at 75\,dB, which corresponds to $\sim$2000 photoelectrons.

The dynamic range of the test pulse is 0.3 to 5000 photoelectrons, which fully covers the dynamic range of the readout system. This is very useful for the quality control of the Dragon board before the installation and also for the performance check after the installation.

\section{Summary}
With all the functionalities of the SCB described above it will be possible to operate the modules of the camera of the LST in a robust way.
Through the SCB steering and monitoring of the PMTs will be realized.
A characterization of the modules was done with the help of the pulse injection functionality.
In the same way, performances can be checked  and controlled even after installation of PMT modules in the camera.

\acknowledgments
We gratefully acknowledge support from the agencies and organizations listed under Funding Agencies at this website: http://www.cta-observatory.org/.
We also acknowledge the great help from Open-It Consortium on the hardware development of the Dragon board.


\begin{thebibliography}{99}
\bibitem{abc}LST team for the CTA consortium, \emph{Large Size Telescope Technical Design Report}, LST-TDR/140408, v. 2.1, 03 May 2015, available at http://cta-observatory.org/

\bibitem{shu}S. Masuda et al. for the LST team, \emph{Development of the photomultiplier tube readout system for the first Large-Sized Telescope of the Cherenkov Telescope Array}, these proceedings, ID-862

\end{thebibliography}
\end{document}